# Optimized periodic control of chaotic systems

Robert Mettin

*Institut für Angewandte Physik, TH Darmstadt, Schloßgartenstraße 7, D-64289 Darmstadt, Germany* [1]

Thomas Kurz

*Drittes Physikalisches Institut, Universität Göttingen, Bürgerstraße 42–44, D-37073 Göttingen, Germany* [2]

**Abstract**

In this work, we demonstrate the open-loop control of chaotic systems by means of optimized periodic signals. The use of such signals enables us to reduce control power significantly in comparison to simple harmonic perturbations. It is found that the stabilized periodic dynamics can be changed by small, specific alterations of the control signal. Thus, low power switching between different periodic states can be achieved without feedback. The robustness of the proposed control method against noise is discussed.

## 1 Introduction

Recently, many ideas have been proposed to influence chaotic systems according to some desired goal dynamics. Besides a large variety of feedback control techniques (see e.g. [1–4] and references therein), several methods that dispense with system state measurements have been proposed. Such non-feedback or open-loop techniques are the only way of control in situations where the system state is not immediately accessible or when the dynamics is so fast that computation and injection of the control signal cannot keep pace with it.

One can distinguish essentially two main approaches to open-loop control of chaos. *Entrainment control methods* (see e.g. [5–9]) use system model equations and specify goal dynamics to construct control forces. If the desired motion is periodic, the resulting forces are periodic, too, but may have quite a large amplitude and a complicated shape. Typically, complete entrainment to the goal dynamics requires as many control forces as there are dimensions of the system.

[1] E-mail: Robert.Mettin@nlp.physik.th-darmstadt.de
[2] E-mail: tkurz@physik3.gwdg.de

In contrast, many examples have been given in the literature where chaos is suppressed and periodic motion is established by exposing a system to only one, quite simple periodic signal or parameter modulation [10–19]. We want to refer to these methods as *periodic perturbation methods*. While such methods may work with small forces, they cannot achieve an arbitrary goal dynamics; typically, the controlled periodic orbit closely traces a part of the unperturbed motion. This part may be an unstable periodic orbit embedded in the chaotic attractor or a section of a chaotic trajectory.

In most cases, periodic perturbation methods utilize harmonic forces [10–14,19] with amplitude and frequency as parameters. Some authors add a second harmonic force [15–17] or use one-parameter families of anharmonic drivers [18], leading to more advantageous effects. However, a detailed consideration of non-harmonic periodic control forces, characterized by many parameters and thus offering many degrees of freedom, is still lacking.

As an extension to previous work on periodic perturbation methods, in this article the application of multi-parameter periodic perturbations for the control of chaotic systems is investigated. Our results show that with this method the necessary control power, compared to harmonic driving, is significantly decreased. Furthermore, we show that it is possible to switch between different periodic system states. As pointed out by Ott *et al.* [1], it is a specific property of controlled chaos to be driven to different "macroscopic" states by "microscopic" control signal changes. Here, we demonstrate this feature for control without feedback.

Our approach is to optimize parameters of a periodic control signal with respect to a goal demand, given in terms of a scalar cost functional (which is a usual procedure of optimal control theory[20]). The control signal is described by a finite set of real parameters (e.g., Fourier modes) and thus is restricted to a certain control function space. The quality of an arbitrary periodic control signal from this space is evaluated in terms of the cost functional, which is to be minimized. This concept allows a certain flexibility in choosing the aim of control, since the cost functional can measure the performance as well as other properties of the signal. For instance, we could demand the suppression of chaos without being attached to a specific goal trajectory, or with certain constraints being satisfied. When dealing with chaotic systems, however, performance evaluations may be expensive (e.g., the calculation of Lyapunov exponents), and the "cost landscapes" may be rugged or even "fuzzy". Thus the cost functional has to be chosen carefully, with a trade-off between calculational effort and desired characteristics of the goal dynamics. Note that the numerical optimization procedure, for fixed system parameters, is carried out only once in the design phase of the control signal, and no further real-time calculations are needed with this method. Therefore, a large computational effort may be tolerable to achieve an optimal design.

## 2 Optimization process

The problem of constructing a suitable control function is formulated as a high-dimensional optimization process. We assume that the equations of motion of the system and the influence of the control are known. A general formulation for a dynamical system in $\mathbb{R}^m$ governed by an ordinary differential equation reads

$$\dot{\mathbf{x}} = \mathbf{v}(\mathbf{x}, \mathbf{u}(t)), \tag{1}$$

where $\mathbf{x} \in \mathbb{R}^m$ is the system state, $\mathbf{u} \in \mathbb{R}^s$ is the control signal, and $\mathbf{v} : \mathbb{R}^{m+s} \to \mathbb{R}^m$ is the vector field, which is assumed not to depend on time explicitly.

At first, we have to choose the dimension $d$ of control parameter space and a base of periodic functions $\{\Phi_n\}$, in which the control signal $\mathbf{u}(t)$ is represented:

$$\mathbf{u}(t) = \sum_{n=1}^{N} \alpha_n \Phi_n(t), \tag{2}$$

$$\Phi_n(t) = \Phi_n(t+T). \tag{3}$$

The expansion coefficients $\alpha_n$ and possibly the control period $T$ are subject to the optimization process (thus $d = N$ or $d = N+1$). As we have a finite series up to $\Phi_N$, the space of possible control functions is restricted, and an optimal control can only be given with respect to this space. When $N$ is increased, optimized solutions found for $N' < N$ are still contained in the enlarged control space. In order to obtain the best as possible control signal it would thus seem desirable to choose a large $N$. However, too large a dimension of the search space may significantly decrease performance of the optimization procedure. This problem can partially be avoided by a proper selection of the basis functions.

Next, a suitable cost functional has to be defined. We consider general cost functionals which depend on the coefficients $\alpha_n$ as well as on sample trajectories $\Gamma_i(t)$ of the controlled system:

$$\mathit{cost} = f\left[T, \{\alpha_n\}, \{\Gamma_i(t)\}\right]. \tag{4}$$

Each trajectory $\Gamma_i(t)$ is defined by its start and end time, $t_{i0}$ and $t_{i1}$, and its starting point $\mathbf{x}_{i0}$. Implicitely, it also depends on the $\{\alpha_n\}$, i.e., on the control. If the system is deterministic, and if $t_{i0}$, $t_{i1}$ and $\mathbf{x}_{i0}$ are fixed, a well-defined cost landscape exists in the $d$–dimensional control parameter space. By the parameters of the control, $\alpha_n$ (and possibly $T$), the parameter space of the original system is enlarged considerably. Typically, the dynamics will exhibit a rich and extremely complicated bifurcation structure in certain parts of this augmented space, as is well known even for quite simple systems (e.g., nonlinear oscillators) with just a few parameters [21,22]. In particular, for a chaotic system it can be expected that windows of periodic motion will be close to the chaotic state and can be reached by a suitable small (control) parameter change. Thus, the task of optimization here is to find such a change satisfying certain constraints, for instance, to yield minimum power of the control. It is clear, then, that the optimization operates in a high-dimensional space occupied by a complicated bifurcation structure with fine details and self-similar features, leading to rugged or non-smooth cost landscapes. If the system investigated is additionally subject to random perturbations (noise), fine details of the bifurcation structure are smeared out, possibly yielding a smoother, but "fuzzy" cost function. The same is the case if one operates with randomly selected trajectory segments $\Gamma_i$ to eliminate the dependence of the cost functional on arbitrarily preselected initial conditions $\mathbf{x}_{i0}$. For these reasons, one should draw attention to the choice of a suitable optimization procedure. In the examples to be presented, we used an optimization algorithm ("amebsa" of Ref. [23]), which is a combination of the simulated annealing technique and the downhill simplex method. Other methods of optimization (e.g., genetic algorithms) are currently under investigation. Due to the high dimension of the problem, a *direct* search for the optimum, however, e.g. by bifurcation continuation techniques [24], seems presently not feasible. Thus one has to rely on stochastically guided search methods.

## 3 Examples

As a first example, we perform periodic control of a Rössler system [25]. A control force which acts additively on the first vector field coordinate is applied:

$$\begin{aligned} \dot{x}_1 &= -x_2 - x_3 + u(t) \\ \dot{x}_2 &= x_1 + a x_2 \\ \dot{x}_3 &= b + (x_1 - c)x_3 \end{aligned} \tag{5}$$

Parameters are set to $a = b = 0.2, c = 4.6$ which, without control, lead to chaotic behavior. The control signal is represented by a Fourier series up to the 5th order with zero mean:

$$u(t) = \sum_{n=1}^{5} a_n \cos(n\omega t) + b_n \sin(n\omega t). \tag{6}$$

Because the uncontrolled system is autonomous, we have the freedom to choose the phase of the periodic control. This is done by always setting $b_1 = 0$. To avoid frequency "run-away", $\omega$ is bound to the interval $[0.520, 0.562]$, which is centered at a strong component of the Fourier spectrum of the uncontrolled system. The cost of a control force is defined in the following way: If, after transients have died out, no periodicity (up to period 8) is recognized in the controlled system, we assign to *cost* a very large value (100, say). Otherwise, *cost* is set to the control signal power:

$$P = \int_0^T |u(t)|^2 dt = \frac{1}{2} \sum_{n=1}^{5} a_n^2 + b_n^2 \quad , \quad T = \frac{2\pi}{\omega}. \tag{7}$$

Thus, suppression of chaos should be achieved with minimum control power if the cost functional is minimized. In absence of an analytic criterion, periodicity is detected by calculation of one or more trajectories under control of the signal to be assessed. Rejecting transients, mean recurrence distances $\bar{d}_k$ after $k$ control periods are determined for this purpose:

$$\bar{d}_k = \frac{1}{M} \sum_{m=0}^{M-1} |\mathbf{x}((m+k)T) - \mathbf{x}(mT)| \quad . \tag{8}$$

If $\bar{d}_{min} = \min_k\{\bar{d}_k\}$ is smaller than a threshold $\epsilon$, the controlled system is supposed to be periodic. As pointed out before, the result of the calculation may depend on the initial conditions of the trajectory. In the numerical examples, we used $k = 1, \ldots 8$, $M = 50$, $\epsilon = 0.01$, and a single trajectory with fixed initial conditions, $\mathbf{x}(t{=}0) = (5, 5, 5)$.

To start optimization, we take a sinusoidal signal with large amplitude that suppresses chaos. For this one-mode signal, frequency and amplitude are optimized, giving a signal of lower power that leads to periodicity. In the next step, we optimize a two-mode signal, starting with the result of the one-mode optimization. Successively, the number of modes is increased up to five with the optimization being repeated each time with the result of the previous step. The dimension of the problem increases by two with each additional mode, so the full problem with five Fourier modes is ten-dimensional.

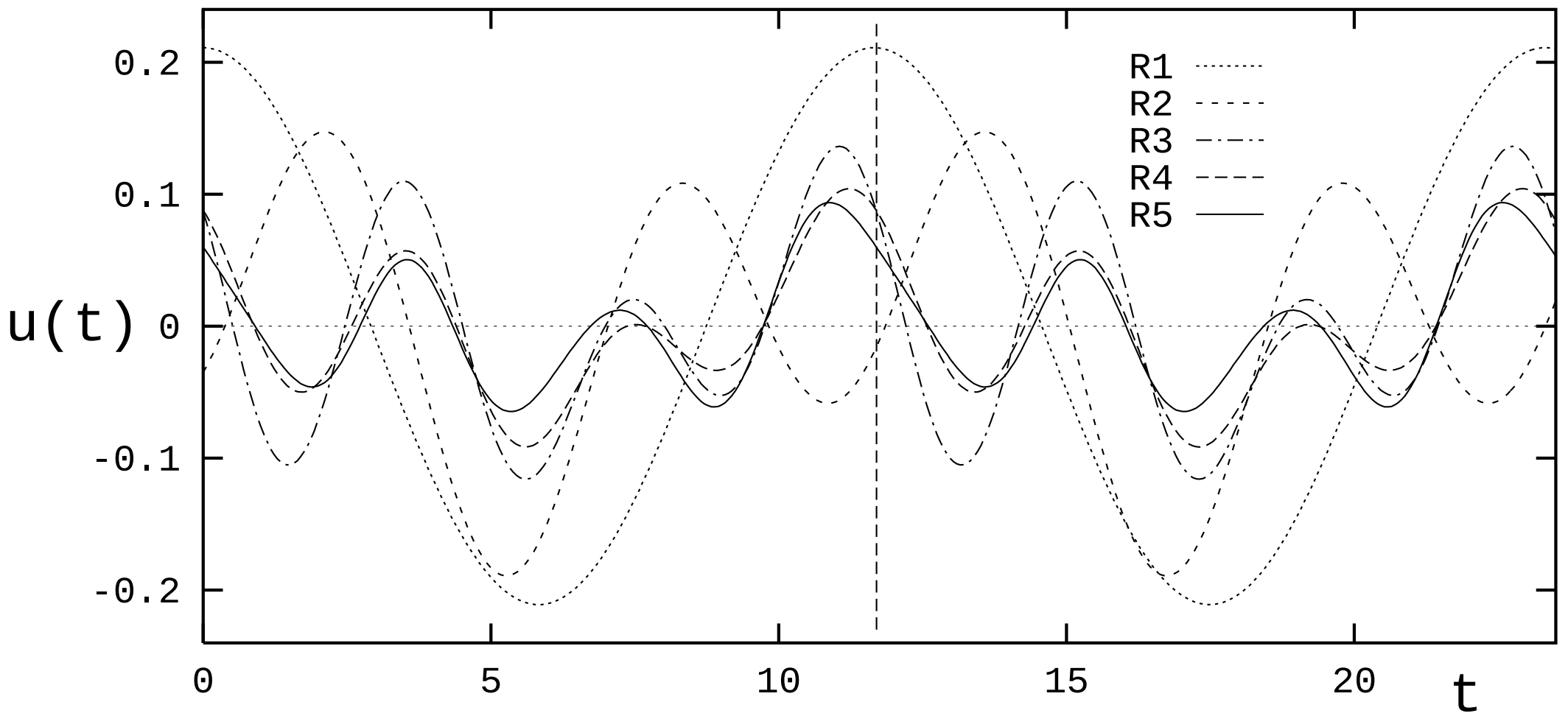

Fig. 1. *Time plots of the best control signals found for the Rössler example. The signal denoted by* R$n$ *is a result of the optimization with* $n$ *Fourier modes. The periods of all signals vary slightly around* $T = 11.7$*, which is marked by the vertical dashed line.*

The best control signals found in successive optimization steps for the Rössler system are shown in Fig. 1, denoted with labels R1 to R5. Their periods are only slightly different and are approximately indicated by the dashed vertical line. In this example, the signal shapes seem to converge to some limiting shape. The signal power is decreasing monotonically, as can be seen in Table I. Fourier coefficients and frequency of the best signal, R5, are shown in Table II. Note that this signal is *optimized*, not *optimal*. There is a certain chance that five-mode control signals of lower cost may exist since the control function space is only sparsely probed by the optimization algorithm. However, the signals were calculated with an acceptable effort. Also, the algorithm rarely steps into too small, isolated "valleys" of the cost landscape (which might contain better signals) and thus provides a certain robustness against noise.

| **#modes** | **signal** | **power** | **signal** | **power** |
|---|---|---|---|---|
| 1 | R1 | $2.23\times10^{-2}$ | T1 | $7.14\times10^{-6}$ |
| 2 | R2 | $9.95\times10^{-3}$ | T2 | $1.09\times10^{-7}$ |
| 3 | R3 | $5.42\times10^{-3}$ | T3 | $6.35\times10^{-8}$ |
| 4 | R4 | $2.69\times10^{-3}$ | T4 | $3.81\times10^{-8}$ |
| 5 | R5 | $2.04\times10^{-3}$ | T5 | $2.46\times10^{-8}$ |

Table I. *Power of the optimized periodic signals for control of the Rössler (*R$n$*) and TVA (*T$n$*) system. $n$ denotes the number of Fourier modes used by the optimization algorithm.*

| | **R5** | **R5-2** | **T5** |
|---|---|---|---|
| $\mathbf{a}_1$ | $3.084916\times10^{-2}$ | $6.361023\times10^{-2}$ | $5.638383\times10^{-5}$ |
| $\mathbf{a}_2$ | $1.285033\times10^{-2}$ | $3.130399\times10^{-2}$ | $-1.471160\times10^{-5}$ |
| $\mathbf{b}_2$ | $-1.940677\times10^{-2}$ | $-1.140495\times10^{-2}$ | $1.211148\times10^{-4}$ |
| $\mathbf{a}_3$ | $3.146328\times10^{-2}$ | $5.770271\times10^{-2}$ | $2.111210\times10^{-6}$ |
| $\mathbf{b}_3$ | $-3.797136\times10^{-2}$ | $-4.469001\times10^{-2}$ | $-1.529694\times10^{-4}$ |
| $\mathbf{a}_4$ | $-8.045605\times10^{-3}$ | $-1.114055\times10^{-2}$ | $3.803825\times10^{-5}$ |
| $\mathbf{b}_4$ | $6.721184\times10^{-3}$ | $1.623834\times10^{-2}$ | $8.325271\times10^{-6}$ |
| $\mathbf{a}_5$ | $-7.191359\times10^{-3}$ | $-1.826102\times10^{-2}$ | $-7.358854\times10^{-5}$ |
| $\mathbf{b}_5$ | $1.354798\times10^{-3}$ | $1.610111\times10^{-2}$ | $2.908119\times10^{-5}$ |
| $\omega$ | 0.5372290 | 0.5372290 | 20.76428 |
| **P** | $2.04\times10^{-3}$ | $5.73\times10^{-3}$ | $2.46\times10^{-8}$ |

Table II. *Fourier coefficients ($a_n$, $b_n$), frequency ($\omega$), and power ($P$), according to Eqs. (6) and (7), of the optimized periodic control signals* R5, R5-2 *(Rössler), and* T5 *(TVA).*

There are certain problems connected with the definition of the cost functional. First, the *cost* definition as given above does not assess the degree of stability of the controlled orbit. Furthermore, for the detection of periodicity the recurrence distance method is not always a 'safe' measure. Errors are mainly due to very long transient dynamics, multiple attractors, and band structured or intermittent chaos.

All this turned up in the calculations, and thus we used the recurrence distance calculation with rather tight parameters. Therefore, the procedure has probably discarded some good control forces. The same difficulties also appeared when using other, quite different cost functionals, including local divergence rates for instance.

Table I shows that the cost functional in fact decreases when the number of Fourier modes is increased. The gain of an additional mode, however, gets smaller and smaller, which is (besides the calculational effort for the optimization process in higher dimensions) an *a posteriori* argument for limiting the mode number.

In the next example, quite the same procedure was used to regularize the behavior of a model of a chaotic chemical oscillator. The system under consideration is known in the literature as the Three-Variable Autocatalator (TVA). Using a parametric control force in this case, the equations of motion read

$$\begin{aligned} \dot{x}_1 &= \mu[1 + u(t)](\kappa + x_3) - x_1(1 + x_2^2) \\ \sigma\dot{x}_2 &= x_1(1 + x_2^2) - x_2 \\ \delta\dot{x}_3 &= x_2 - x_3 \quad . \end{aligned} \tag{9}$$

The uncontrolled system is chaotic for $\mu = 0.154$, $\kappa = 65.0$, $\sigma = 0.005$, and $\delta = 0.02$ (see [26]), which are the parameters used in our calculations. The optimization was carried out as in the Roessler example, except that the control signal frequency was restricted to the interval $[18.0, 22.0]$, and that the initial conditions were chosen as $\mathbf{x}(t{=}0) = (0.6, 15.0, 17.0)$. Results of the optimization are shown in Table I, where the signal powers of the best control signals found, labeled T1 to T5, are given. In Fig. 2, we show the signals T2 to T5 vs. time (the sinusoidal signal T1 with the amplitude $a_1 \approx 0.00378$ is not shown).

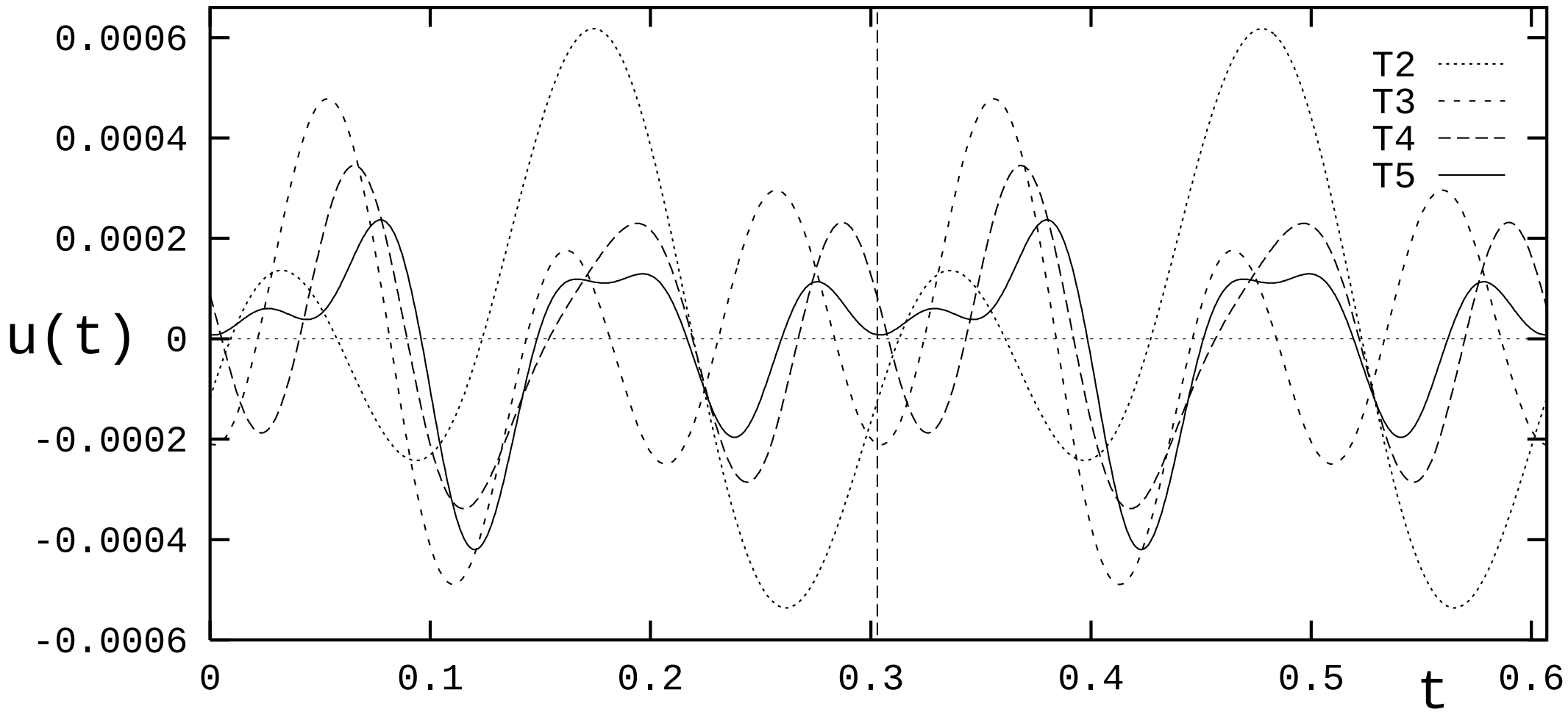

Fig. 2. *Time plots of the best control signals found for the TVA example. The signal denoted by* T*n is a result of the optimization with n Fourier modes (the best one-mode signal is excluded because of its large amplitude). The periods of all signals vary slightly around* $T = 0.303$, *which is marked by the vertical dashed line.*

Again, the vertical dashed line marks the approximate period of all signals. The Fourier coefficients and frequency of the best five-mode signal, T5, are given in Table II. Albeit parametric actions cannot be well assigned a "power" to be compared with the system power, we nevertheless used this quantity for minimization. Again, we achieve a significant reduction of the system perturbation needed for regularization (see Table I). Note that the modulated parameter $\mu$ does not leave the region where the uncontrolled TVA behaves chaotically.

## 4 Switching of control

To demonstrate the ability of switching by open-loop control, we return to the Rössler example. Besides the optimized five-mode control signal R5 of Fig. 1 which stabilizes a period 3, we have optimized another five-mode signal with the same frequency, but with the constraint to stabilize a period 2. Fourier coefficients and power of this signal, which we denote by R5-2, are presented in Table II. Figure 3 shows both signals vs. time.

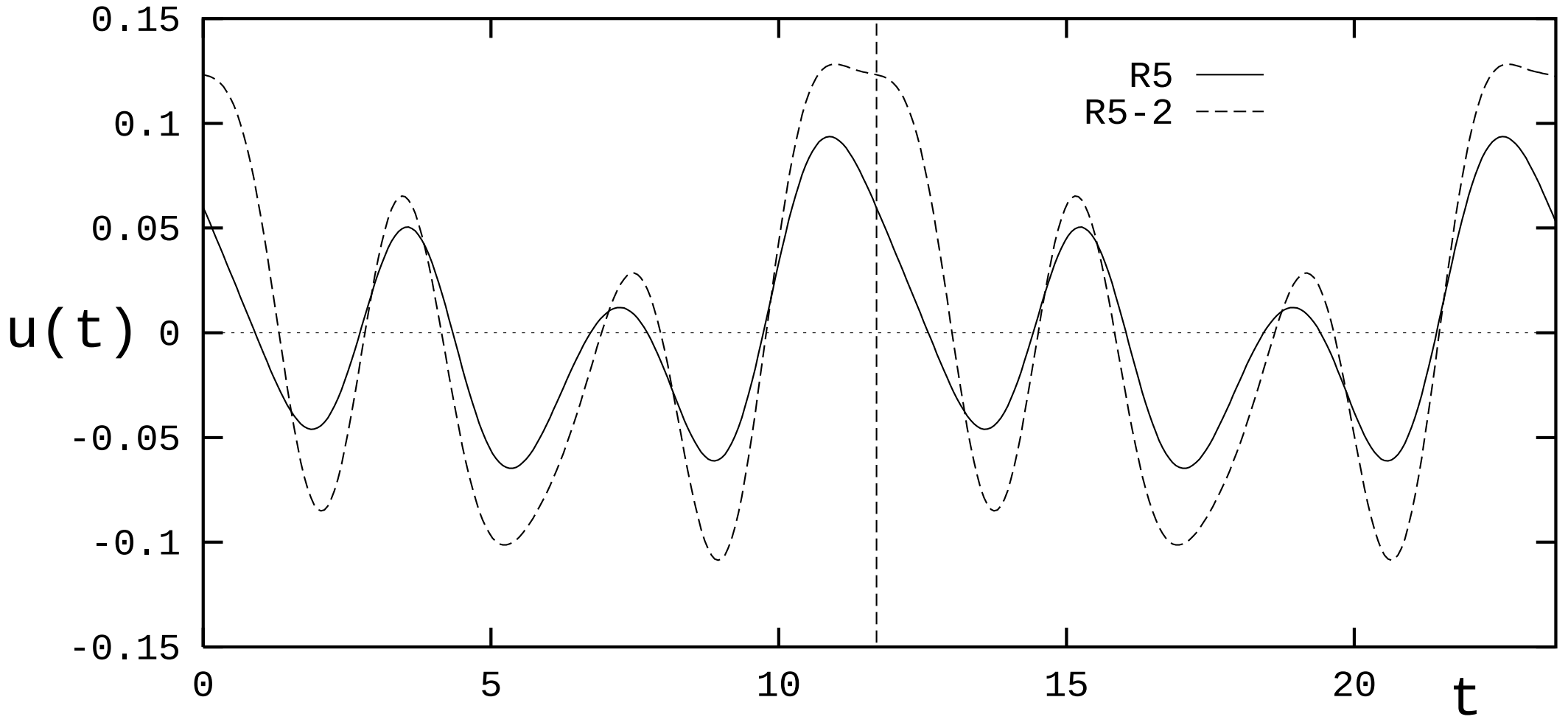

Fig. 3. *Time plots of the two control signals used in the switching of the Rössler system. Both include five Fourier modes and have the same period* $T = 11.69554$*. The signal* R5 *is identical with the five-mode signal* R5 *shown in Fig. 1.*

By switching between the two control signals the chaotic Rössler system Eq. (5) can be forced alternatively to a period-2 or to a period-3 state. This is shown in Fig. 4a, where the first coordinate of a Poincaré section is plotted vs. time. (Since the trajectories wind around twice in one control period, the different states appear as period-4 and period-6). Control is active for $t = 5000$ to $t = 45000$, and the signal is switched between R5 and R5-2 every 5000 time units. Here, more or less long transients appear until the controlled system settles down onto the desired periodic orbit. The transient times range from only a few up to over 300 control signal periods. However, transient times can be minimized by switching at selected phases. The length of a transient depends on the system state at the moment when the switching takes place, and on the phase of the control signal that is turned on. For illustration, consider the case of switching from the signal R5 to the signal R5-2. If we assume that the system state, at the time of switching, is on the R5-controlled trajectory, then the actual state is given, up to a possible ambiguity due to a period $> 1$, by the phase of R5. For all such phases of R5 (and corresponding system states), we can determine that initial phase of the signal R5-2 which yields the shortest transient. The best pair of ending and initial phases is given by the overall minimum. Thus, by suitably selecting switching times and initial phases, optimized-phase switching can be achieved which is illustrated in Fig. 4b for the example considered before. Note that due to the phase adjustment the time difference between successive switching events is not fixed but may deviate from the interval used in Fig. 4a (5000 time units) by up to one driver period $T \approx 11.7$.

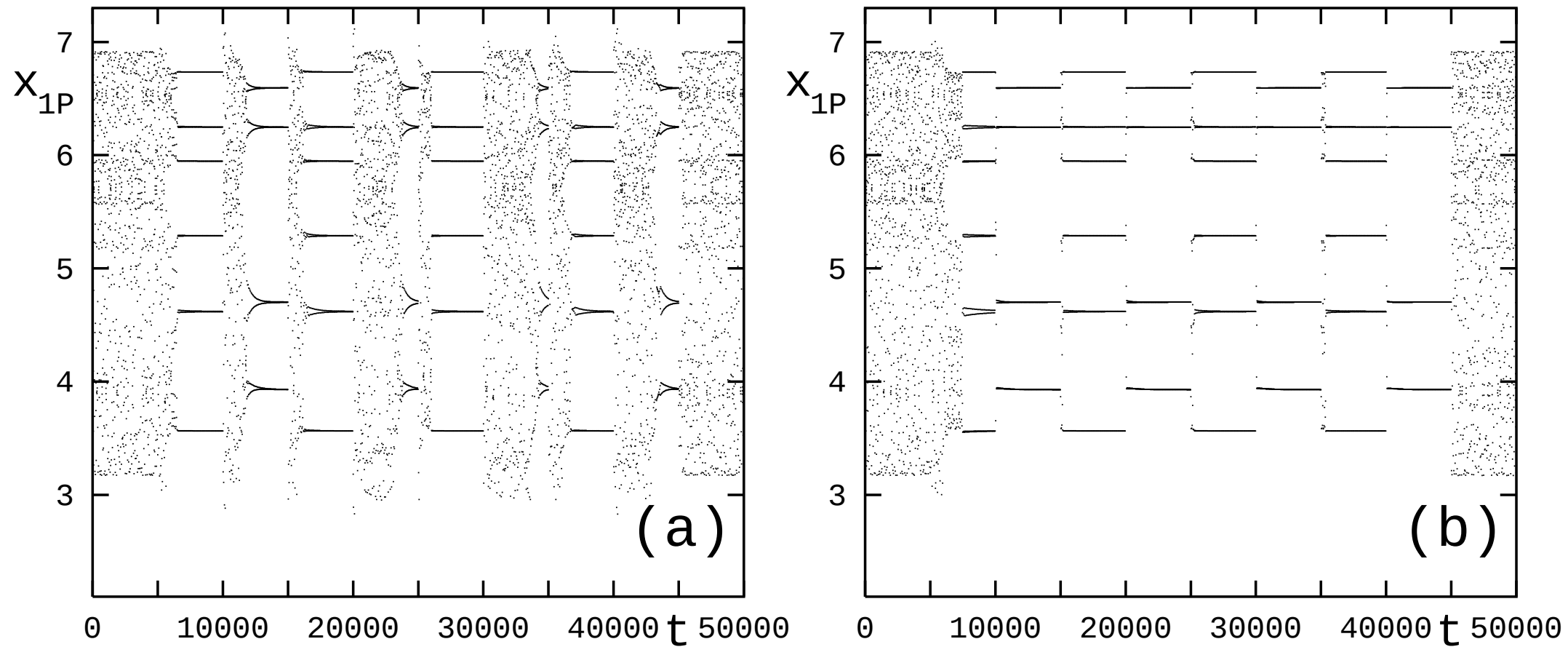

Fig. 4. *Switching of periodic control of the Rössler system: first coordinate* $x_{1\mathrm{P}}$ *of a Poincaré section vs. time t. Control is turned on at* $t = 5000$ *and turned off at* $t = 45000$*. Switching takes place every 5000 time units. Arbitrary relative phases of the control signals give rise to possibly long transients (a). These can be reduced by selecting optimized switching phases, except for the turn-on transient (b).*

This result demonstrates that the system behavior can be altered merely by a slight but carefully chosen change of the control signal shape, at constant frequency. (More generally, the switched control signals could also have different frequencies.) Note that the maximum amplitude of both signals is about 1% of the maximum of $|\dot{x}_1|$.

## 5 Control of noisy systems

Finally, we briefly address the robustness of optimized periodic control against noise. To this end, Gaussian white noise is added to the system by the Box-Mueller algorithm [27] which is applied to all coordinates. The noise level is given by the variance $\sigma$ of the Gaussian distribution. Performance of control can still be measured by mean recurrence distances $\bar{d}_k$, since one expects the control to stabilize a noisy periodic orbit instead of an exact one. The minimum mean recurrence distance $\bar{d}_{min}$, however, increases with the noise level.

For the control signals R5 and R5-2 of the previous example (Rössler system), Figs. 5a and 5b show $\bar{d}_{min}$ vs. the variance $\sigma$ of the applied noise. Here, recur-rence distances were averaged over 1000 periods. Both graphs exhibit a roughly linear part at lower noise where fluctuations of the averaged values $\bar{d}_{min}(\sigma)$ are also low. At noise levels larger than $\sigma \approx 0.00012$ (for R5) and $\sigma \approx 0.0015$ (for R5-2) the fluctuations increase significantly, and $\bar{d}_{min}$ finally reaches values that are found for the uncontrolled system ($\bar{d}_{min} \approx 1.5$); that is, periodicity is completely destroyed by the noise. Note that this noise level threshold is much larger for the control signal R5-2 which has larger power. Also, the slope of the linearly increasing part of the graph is smaller than that of R5, which can be seen from the insets of Figs. 5a and 5b. We suspect that robustness against noise simply increases with control power. Indeed, the slope of the linear part of $\bar{d}_{min}(\sigma)$ at low $\sigma$ seems to scale reciprocally with the power of the control signal. Thus we tentatively conclude that the better a control signal is optimized (with respect to minimum power), the less robust it is (with respect to Gaussian white noise in the controlled system). Qualitatively, this may be explained by features of the bifurcation structure: regions of periodic motion in parameter space usually are confined by bifurcation hypersurfaces with cusp-shaped structures that appear as stripes or tongues in two-dimensional parameter sections. Control signals with lowest power are located at the very narrow tips of such tongues. Then, even small noise perturbations result in small effective parameter shifts out of the periodic region. As a consequence, for a noisy system one should incorporate into the definition of the cost functional a trade-off between power reduction and stability of the controlled motion.

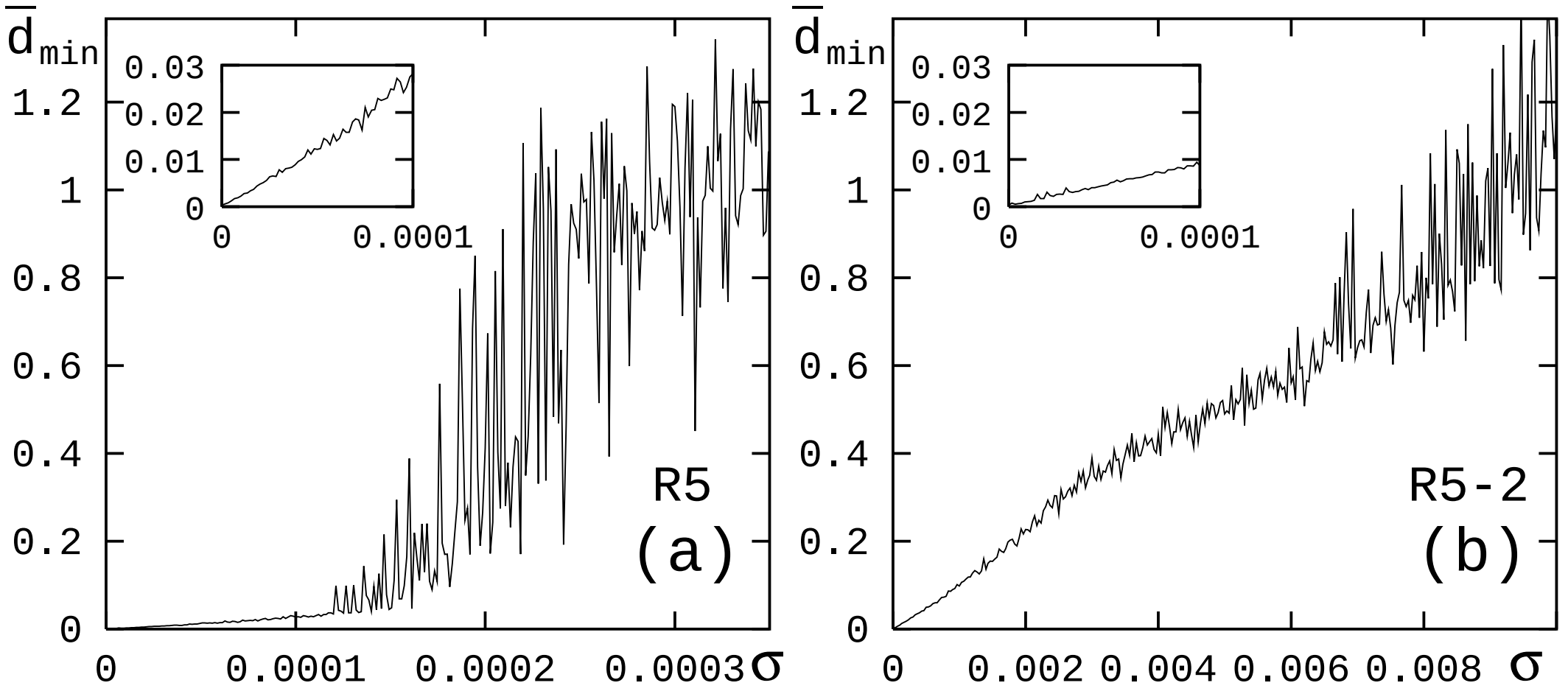

Fig. 5. *Influence of Gaussian noise on the controlled system: mean recurrence distance* $\bar{d}_{min}$ *vs. noise variance* $\sigma$ *for applied control signals* R5 *(a) and* R5-2 *(b). Note the different scales of the abscissa. The insets show magnifications of the low noise parts with the same scale for (a) and (b), respectively.*

## 6 Summary and discussion

In conclusion, we have shown that it can be advantageous to use multiple-mode signals for periodic control of chaotic systems. To determine such signals, a cost functional has been defined which is minimized by a high-dimensional optimization algorithm. The functional contains the control signal power as well as a measure of control performance, i.e., mean recurrence distances. We have demonstrated regularization of chaotic motion by very low control power for the Rössler system (additive control) and the TVA system (parametric control). Different shapes of the perturbing signal can effect different periodic states, offering the possibility of switching the system by very low effort. Switching transient times can be drastically reduced when adjusting the phases of the switched signals, as has been demonstrated for the Rössler system. Furthermore, the robustness of optimized periodic control against system noise has been investigated. The results suggest that a reduction of control signal power is partially accompanied by a higher sensitivity with respect to noise. Nevertheless, periodic perturbation controls can be made more effective if higher modes or other available basis functions are added to harmonic control forces (compare [18,28]).

The method introduced in this paper may be extended in several ways. For example, the representation of the periodic control signal is not restricted to a Fourier base but may be given in any other suitable function system (e.g., step functions). Also, the cost functional may be defined in such a way as to achieve more general regularization goals. This is especially important in systems with a large number of degrees of freedom, or when the goal dynamics is not easily defined in terms of a certain goal trajectory (which would be an application for model-based control [5–9]), e.g. when Lyapunov exponents are of interest. Finally, the performance of optimization may be improved by a proper selection of the algorithm. Research on optimization methods has been very innovative in recent years, and further impulses and improvements are to be expected. All these issues are currently under investigation; the results will be reported in a forthcoming publication[29].

## Acknowledgement

We thank W. Lauterborn, J. Holzfuss, U. Parlitz, and the Nonlinear Dynamics Group at TH Darmstadt for valuable discussions and support. This research was supported by the Deutsche Forschungsgemeinschaft (Sonderforschungsbereich 185).